\documentclass[a4paper]{article}
\usepackage[bottom]{footmisc}
\usepackage{paralist}
\usepackage{amsfonts,amsmath}

\usepackage{tikz}
\usepackage{listings}
\usepackage[backend=bibtex]{biblatex}
\usepackage{dmbiblatex}
\bibliography{Monniaux_completeness_in_static_analysis}
\usepackage{hyperref}

\newcommand{\ve}[1]{\mathbf{#1}}

\title{Completeness in Static Analysis by Abstract Interpretation, a Personal Point of View}
\author{David Monniaux}

\begin{document}
\maketitle


\abstract{%
Static analysis by abstract interpretation is generally designed to be “sound”, that is, it should not claim to establish properties that do not hold—in other words, not provide “false negatives” about possible bugs. A rarer requirement is that it should be “complete”, meaning that it should be able to infer certain properties if they hold. This paper describes a number of practical issues and questions related to completeness that I have come across over the years.

}

\section{Introduction}

The concept of \emph{completeness}, under several definitions, permeates mathematical logic. A proof system is deemed complete with respect to a semantics if it can prove all properties that are true in this semantics. 
For instance, Gödel's completeness theorem states that first-order logic (that is, any reasonable proof system for it) can prove any property true in all models.
A decision procedure for a logic (that is, a procedure that answers ``true'' or ``false'' for any formula in the logic $F$) is expected to be \emph{sound} (it does not declare to be true formulas that are not true) and \emph{complete} (it declares to be true all formulas that are true).

The concepts of completeness
\begin{inparaenum}[(a)]
\item for a proof system with respect to a class of properties
\item for a procedure that searches for proofs of such properties within such a proof system
\end{inparaenum}
are related, but distinct. Obviously, if a proof system is incomplete, then so is any procedure that searches for proofs within that system. However, it is possible to have a complete proof system and an incomplete procedure for searching within it, for instance for practical efficiency reasons.

In abstract interpretation, completeness, at least at the global level (proving properties of whole programs), is often forgone straight from the start when designing the abstraction (for instance, interval arithmetic is used even though it is obvious that it cannot prove safety in general, even if the property to prove is itself an interval), and there thus may be little reluctance to adding further incompleteness in the solving procedure by using approaches such as widening operators~\cite{CousotCousot77,DBLP:journals/logcom/CousotC92}.
Yet, it is interesting to control this further incompleteness, both from a theoretical and a practical points of view.
Completeness in the solving method ensures some predictability in the analysis outcome, while the brittleness of some incomplete approaches (the approach succeeds or fails depending on seemingly inconsequential aspects of the program and the property) surprises end users: for instance, providing more precise information on the precondition of a program may prevent the analysis procedure from proving a property that it could prove with a less precise precondition.

In this paper, I discuss various instances of completeness and incompleteness that I have came across, without any pretense of thorough theoretical discussion and, ironically, no pretense of completeness.
For more theoretical discussion, Giacobazzi et al. \cite[Sec.~7]{DBLP:journals/jacm/GiacobazziRS00} surveyed the literature extensively.
For a study of completeness with respect to the relationship between the semantics and the abstract domain, and constructions to make the domain complete, see \cite{DBLP:conf/amast/GiacobazziR97}.

\section{Completeness of the Abstraction: the Case of LRU Caches}\label{sec:cache}
It is well-known that it is equivalent to
\begin{inparaenum}[(a)]
  \item
    evaluate an integer multivariate polynomial over integer inputs, then take the final result modulo $N$
  \item
    do the same computation but reducing modulo $N$ at every step.%
  \end{inparaenum}%
\footnote{%
For $N=9$, this is the basis for the method of ``casting out nines'', named so because reduction modulo $N$ can be implemented over numbers in decimal form by summing the digits of the number, with $9$ being replaced by $0$ (repeat the process until the result consists on one single digit).
Schoolchildren used to apply this method to check their hand computations: the final result of an error-prone integer computation, taken modulo $9$, should be identical to the result obtained with reduction at every step. This is of course not a sound check: if results differ, one is sure that there has been some error somewhere (no false positives), but they can coincide by chance.}
Due to a strong mathematical property (a ring morphism), it is possible to replace an expensive computation, possibly involving large numbers, by a simpler one that abstracts the expensive one by operating on small abstractions of the data.

This example is ideal, and one could doubt the existence of complete abstractions that significantly simplify computations for nontrivial program analysis problems; yet we came across one such example, where we perform an exact analysis defined by a composition of exact abstractions and an exact solving procedure.

All current processors use some form of cache memory: frequently accessed code or data is retained in fast memory close to the processor core, avoiding much slower accesses to main memory.
Cache analysis consists in determining, given a program and a description of a processor cache subsystem, which memory accesses are always ``cache hits'' (data already in cache), which are always ``cache misses'' (data not in cache, thus access to external memory), or more complex properties (``the first time this statement is executed, the access is a cache miss, then subsequent accesses are hits'').
Such an analysis is for instance used as a prerequisite for worst-case execution time analysis.

A possible approach to solving the cache analysis problem would be to apply a model-checker to the combination of the program under analysis (or some suitably simplified model thereof) and of the cache system.
A first and obvious abstraction is, for the cache system, to retain only the addresses of the memory blocks currently stored within the system, but not their contents.
This abstraction is exact: the cache system finds data, or decides to evict data to make room for new contents, according to addresses only, never according to contents.
The resulting model of the cache system, for a fixed processor model, is finite: the cache consists of a fixed number of cache lines, buffers, etc., labeled with addresses taken within a fixed range.
However, such a model is in practice intractable on all but the simplest examples.

The naive vision of cache memory is that of a \emph{fully associative} cache: any cache line may be used to store any block from the main memory, regardless of its address. In reality, a cache is usually split into several \emph{cache sets}, each of which able to retain blocks present only at a given class of addresses in the main memory. The classes of addresses for the different cache sets form a partition of the possible addresses in main memory.
In almost all cases%
\footnote{The exception is the ``pseudo round robin'' cache replacement policy, which involves a counter global to all cache sets. Also, dependencies between cache sets arise if one considers an integrated model of the processor pipeline, pre-fetching units, and caches, because whether or not some data is in some cache set influences indirectly whether some other data will be fetched.}
these cache sets operate completely separately from each other.
It is thus possible to perform cache analysis separately for each cache set; this is again an exact abstraction.
However, this still results in intractable models.

Each cache set is composed of a fixed number of \emph{cache lines}; the number $N$ of lines is also known as the number of \emph{ways} or the \emph{associativity} of the cache.
Each cache line may be empty or nonempty; a nonempty cache line stores a block from main memory: its address and its contents (we have already seen that the contents are not relevant for analysis).
For simplicity, we shall consider here the case of single-level, read-only caches.
A read from a memory location not currently in the cache triggers a load from main memory into the cache. Except in the rare case when the relevant cache set still has empty lines, this involves evicting a block from the cache set.
The selection of the block to be evicted is made according to a \emph{cache replacement policy}.
The most obvious replacement policy is to evict the \emph{least recently used} (LRU) block.%
\footnote{Some processors implement this LRU policy. It is however more common to implement so-called pseudo-LRU policies, meant to provide the same practical performance at a fraction of the hardware cost~\cite{DBLP:journals/pieee/HeckmannLTW03,Gille_2007}. Unfortunately, these pseudo-LRU policies are very different from the point of view of cache analysis~\cite{DBLP:journals/pieee/HeckmannLTW03,Monniaux_FMSD22,Monniaux:2019:CCA:3368192.3366018}.}
A cache set is therefore modeled as a sequence of at most $N$ block addresses, in increasing \emph{age}: the \emph{youngest} block is the one accessed most recently, the \emph{oldest} is the least recently accessed.
The positions of blocks within the set change as data is accessed.

Let us take an example with $N=4$. Assume $a,b,c,\dots$ denote distinct addresses in memory, and assume the cache set initially contains $abcd$, meaning that $a$ is the youngest block and $d$ the oldest.
If the processor requires block $d$, then it is rejuvenated and the cache set then contains $dabc$.
If then the processor requires block $e$, then the oldest block ($c$) is evicted and the cache set contains $edab$.

Let us now see our approach to analyzing programs over LRU caches~\cite{Touzeau:2019:FEA:3302515.3290367}.
Consider the case where we want to know whether a certain fixed block $a$ is within the cache at certain program locations.
Then, the blocks older than $a$ in the cache are unimportant;
so is the ordering of the blocks younger than $a$.
The idea is that if a block is younger than $a$, then accessing it again will just change the ordering of blocks younger than $a$, but not the age of $a$;
and if it is older, then accessing it will increment the age of~$a$.

A cache set may thus be described, with respect to a block $a$, by an abstract configuration: either ``$a$ is absent'' or ``$a$ is here and the set of blocks younger than $a$ is $\{ \dots \}$''.
Again, this is an exact abstraction.
Cache analysis using that abstraction specialized for accesses to $a$ thus amounts to collecting the sets (for all program locations) of reachable abstract configurations in the cache set where $a$ is to be stored.
This is still very costly.

A collection of abstract configurations for that cache set consists of a Boolean ``is it possible that $a$ is absent'' and a set of sets of size less than $N$ of blocks distinct from~$a$.
A crucial observation is that, in this set of sets, only the minimal and maximal elements (with respect to the inclusion ordering) matter:
if it is possible that $a$ is preceded by $\{b\}$, $\{b,c\}$, $\{b,c,d\}$, then, for checking if it is possible that $a$ (at some later point) is out of the cache, it is only necessary to retain $\{b,c,d\}$, which subsumes the other two cases;
and for checking if it is possible that $a$ (at some later point) is in the cache, it is only necessary to retain $\{b\}$, which again subsumes the other two cases.
The reason for this subsumption is that if a sequence of steps leads from a state where $a$ is preceded in the cache by a set $P$ of blocks to the next access to $a$, and that access is a miss, then the same sequence of steps of steps, starting with any cache state where the set of blocks younger than $a$ is a superset of $P$, also leads to a miss (\emph{mutatis mutandis} for subsets and hits).
Retaining only the extremal elements is thus, again, an exact abstraction from the point of view of cache analysis.

To summarize, we first gradually simplified the cache state by abstracting away parts and aspects that are irrelevant to the analysis under way, then we simplified the collection of abstract cache states by using a subsumption relation. The resulting abstraction is exact with respect to the properties under analysis.

A this point, what remains is how to implement this abstraction. What is needed is an efficient way to represent sets of incomparable sets of blocks (\emph{antichains}); we opted for zero-suppressed binary decision diagrams (ZDD).
For better efficiency, this analysis, still somewhat costly, is applied to memory accesses only after some cheap approximate analyses \cite{Touzeau_et_al_CAV2017} have failed to classify these accesses.

An advantage of an exact approach is that, since the result is uniquely defined, it is possible to test implementations of various exact abstractions against each other; the results should be identical. This way we discovered a subtle ZDD bug by comparison with a (less scalable) model-checking approach.

Granted, this approach is ``exact'' or ``complete'' only because a simplified model of the program (a control-flow graph, not taking into account the data being processed) is used. This model introduces approximation: for instance, in the following program
\begin{lstlisting}
if (flag) access(a); else access(b);
if (flag) access(a); else access(b);
\end{lstlisting}
the only feasible sequences of accesses are $a,a$ and $b,b$, (and thus in either case the second access is a hit), but an analysis based on control-flow only will not track the value of the flag and consider that sequences $a,b$ and $b,a$ are also feasible, and conclude that the second access may be a miss.
However, tracking the value of data precisely makes the model undecidable.
An intermediate solution that retains decidability is to consider control-flow with precise modeling of procedure calls~\cite{Monniaux_FMSD22}.

On a final note, on many examples~\cite{Touzeau:2019:FEA:3302515.3290367}, worst-case execution time (WCET) analysis ended up being overall faster with the help of this apparently expensive analysis. The reason is that the WCET analysis in use involved enumerating all reachable pipeline states. If the cache analysis is imprecise and cannot conclude about some access, where the exact analysis would conclude to a ``always miss'', this does not normally change the bound on WCET, but this increases the number of pipeline states to consider. In short:
\begin{itemize}
\item incomplete analyses are first applied in order to answer most subproblems exactly; only the subproblems for which the answer is definitely unknown are passed to the more expensive complete analysis;
\item completeness is ensured by a composition of abstractions that do not lose any precision with respect to the properties we are interested in, but which greatly simplify the model to analyze;
\item the resulting model is analyzed exactly;
\item complete analyses have an exactly defined result, thus testing for bugs is easy by comparing results;
\item for the same reason, it is possible to study the complexity of the problem \cite{Monniaux:2019:CCA:3368192.3366018,Monniaux_FMSD22};
\item the extra cost of the complete analysis may be offset by savings in client analyses, because a more precise result means fewer case analyses down the road.
\end{itemize}

\section{Completeness or Incompleteness of the Analysis Method}
A classical question of completeness in abstract interpretation is whether the abstract domain in use is sufficient to prove the properties being sought. Yet, even if the abstract domain is sufficient, it may happen that the abstract interpretation method being used cannot find the necessary invariants within the domain.

\subsection{Widening Operators}
Let us see an example: some simplified version of a dataflow program where \lstinline|i| is an index into some circular array of length 42.

\begin{lstlisting}
i = 0;
while (true) {
  if (trigger()) {
    i = i+1;
    if (i > 42) i = 0;
  }
  assert (i < 1000);
}
\end{lstlisting}

Textbook forward static analysis by intervals \cite{CousotCousot77,Cousot78} will compute, for variable \lstinline|i|, a sequence of intervals $[0,0]$, $[0,1]$, $[0,2]$, \dots, and widen it to $[0,\infty)$.
If narrowing iterations are used, in their simplest form by starting from the $[0,\infty)$ invariant, running the loop body once more and adding the initialization states, one obtains $[0,999]$.
The assertion cannot be proved using that invariant.
Yet the assertion would have been proved if the analysis had guessed the least possible interval, $[0,42]$.
Clearly, the problem is with the inference method (iterations with widening), not the abstract domain (intervals).
Iterations with widening are an \emph{incomplete} method with respect to the abstract domain: they may fail to discover suitable inductive invariants when some exist in the domain.

Note also a surprising characteristic of this incomplete approach: if instead of the precondition $i=0$ we had analyzed the loop with the precondition $0 \leq i \leq 42$, we would be able to prove the assertion correct.
This \emph{non monotonic} behavior (a less precise precondition, or a coarser abstraction of some program semantics leads to more precise analysis results) may surprise end users.

Another surprising characteristic of widening operators for relational domains is that they do not commute, in general, with projection, meaning that adding some extra variables may change the result of the analysis on the original variables even if the extra variables have no influence on the original ones (for instance, if these variables are mere ``observers'').
Consider for instance the program:
\begin{lstlisting}
i=0; j=0;
while (true) {
  if (*) i=0; else {i=1; j=0; }
  if (i==0) j=j+1;
}
\end{lstlisting}
In this program, \lstinline|j| observes the number of last iterations during which \lstinline|i| was 0.
The set of reachable states of this program for $(i,j)$ is $(1,0) \cup \{(0,n) | n \in \mathbb{N}\}$;
in particular the set of reachable values for $i$ is $\{0,1\}$, and it can be computed exactly with interval analysis by postponing widening by one step, or by applying one step of narrowing.
Now consider the sequence of polyhedra, that is, polygons, computed by polyhedral analysis with widening \cite{CousotHalbwachs78}: at step $n$, the polygon is defined by vertices $(0,0)$, $(1,0)$, $(0,n)$.
The lines passing through vertices $(0,0)$ and $(1,0)$ (constraint $j \geq 0$) and the one passing through vertices $(0,0)$ and $(0,n)$ (constraint $i \geq 0$) are stable;
the line passing through vertices $(1,0)$ and $(0,n)$ is unstable and will be discarded by widening.
The resulting system of constraints is $i \geq 0 \land j \geq 0$;
we have lost the $i \leq 1$ constraint obtained by interval analysis.
In some cases, a \emph{stratified} approach, where successive analyses take increasing subsets of variables in consideration, may be able to recoup precision~\cite{DBLP:journals/entcs/MonniauxG12}.

\subsection{Exact Solving}\label{sec:exact}
A safety property is established using forward abstract interpretation by inferring some inductive invariants in the abstract domain and checking that these invariants imply the property.
A complete abstract interpretation method is one that would compute always such invariants if they exist.

Clearly, any method that computes the least inductive invariants in the abstract domain is complete: if there exist invariants in the domain that can prove the safety property, then, a fortiori, the least inductive invariants in the domain can prove that property.

One such method is ascending iterations \emph{without widening} within a domain that satisfies the ascending chain condition (no infinite strictly ascending sequences), which ensures termination. This includes domains of finite height (there is a bound on the length of strictly ascending chains), and in particular finite domains, such as powersets of finite sets.
If $f : L \rightarrow L$ is a monotone function, and $\bot$ is the infimum of $L$, then the sequence $f^n(\bot)$ is ascending and, in all the above cases, becomes stationary. When $f^n(\bot)=f^{n+1}(\bot)$, then $f^n(\bot)$ is the least solution of $x=f(x)$.
The algorithms used in practice are algorithmic improvements over this idea.%
\footnote{In many cases, $L=L_b^m$ where $L_b$ is a base lattice. An element $x$ of $L$ is then decomposed into $x_1,\dots,x_m$, and $f$ is decomposed into its $m$ projections $f_1,\dots,f_m$. The problem is then to find a solution of a system of equations $x_1=f_1(x_1,\dots,x_m), \dots, x_m=f_m(x_1,\dots,x_m)$, typically using a system of working set of variables being updated, with some judiciously chosen ordering for choosing the next update to process.}

An example of an infinite domain of finite height is that of solution sets of systems of linear equations over a fixed number of variables~\cite{DBLP:journals/acta/Karr76}.
When a solution set $S$ is strictly included in a solution set $S'$, the dimension of $S'$ is strictly greater than that of $S$; this dimension cannot be more than~$n$, which thus bounds the height of the lattice.

The domain of intervals has none of these characteristics; but it is however possible to compute the least inductive invariant within that domain of programs such as the one above, by specifying this least inductive invariant as a least fixed point, writing a system of numeric equations that this fixed point should satisfy, and solving this system for the least solution.
Indeed, for the above example, let us write down the equations that $h$ should satisfy for $[0,h]$ to be a fixed point for the loop:

\begin{equation}
  h = \min(\max(\min(42,h+1),h),999)
\end{equation}

Let us solve this equation. In any solution, the outermost ``minimum'' operator must be equal to one of its arguments. Assume it is the second argument, $999$. One can indeed verify that $999$ is a solution to this equation.
Now consider the case where it evaluates to its left argument.
The equation then becomes $h = \max(\min(42,h+1),h)$.
Similarly, the ``maximum'' operator evaluates to one of its operands.
Assume it evaluates to its right argument.
The system then simplifies to $h=h$; but this is under the condition that $h \leq 999$ and $h \geq \min(42,h+1)$.
When $h+1 > 42$, that is, $h \geq 42$, this minimum is equal to $42$;
thus all $42 \leq h \leq 999$ yield valid solutions.
When $h+1 \leq 42$, this minimum is equal to $h+1$; but then the condition becomes $h \geq h+1$, which would imply $h = \infty$, which contradicts $h \leq 999$.
Now assume the ``maximum'' operator evaluates to its left argument.
The system then simplifies to $h = \min(42,h+1)$, which yields only $h=42$ as solution.
We conclude that $h=42$ is the least solution.

The above reasoning by case analysis over the minimum and maximum operators can be automated through exhaustive case analysis or, most subtly, by a SMT (satisfiability modulo theory) solver. Many such solvers can also optimize within the solution space, and thus directly produce the least fixpoint. Alternatively, for proving safety properties, it is sufficient to query the solver for any solution sufficient for proving the safety property.

A more principled approach to the case analysis for the maximum operators is ascending policy iteration \cite{DBLP:conf/esop/GawlitzaS07}, which considers an ascending sequence of fixpoints of systems obtained by picking arguments to the ``maximum'' operators.
All the above approaches generalize to domains other than intervals, for instance to zones~\cite{DBLP:series/natosec/GawlitzaS12}, at the expense of some algorithmic complications.

Let us go further. Assume the abstract domain consists of the sets defined by a parameterized predicate $I(\ve{p},\ve{x})$ where $\ve{p}$ is the vector of parameters, $\ve{x}$ is the program state. Note that this encompasses intervals, octagons, template polyhedra domains, or even polyhedra with a fixed number of faces. Also note that $I$ may contain disjunctions and express non-convex relations.

The inductiveness condition for a transition $\tau_{i,j}$ from program location $i$ to program location $j$ may thus be written as the Horn clause
\begin{equation}
  \forall \ve{x},\ve{x'}~ I(\ve{p}_i,\ve{x}) \wedge \tau_{i,j}(\ve{x},\ve{x'})
  \Rightarrow I(\ve{p}_j,\ve{x})
\end{equation}
Safety conditions may be expressed as
$\forall \ve{x}~ I_i(\ve{p}_i,\ve{x}) \Rightarrow C_i(\ve{x})$, and
program startup may be specified as $\forall \ve{x}~ S_i(\ve{x}) \Rightarrow I_i(\ve{p}_i,\ve{x})$.
Inductive invariants within the abstract domaine suitable for proving the safety properties are thus expressed as the solutions, in the parameters $\ve{p}_i$, of a system of Horn clauses.

If $I$, the transitions $\tau_{i,j}$, the safety conditions $C_i$ and the start conditions $S_i$ are all expressed within a theory admitting algorithmic quantifier elimination, such as Presburger arithmetic or the theory of real closed fields, then the existence of such invariants is decided by quantifier elimination.
For instance, if we consider a program operating over real numbers, and invariants of the form $A(\ve{p}) \ve{x} \leq B(\ve{p})$ where $A(\ve{p})$ and $B(\ve{p})$ have a fixed number $r$ of rows (a polyhedron with at most $r$ faces), then the existence of a suitable invariant is established by quantifier elimination in the theory of real closed fields.
It is even possible to specify that one wants the least inductive invariant and obtain it by quantifier elimination~\cite{Monniaux_POPL09}.

\subsection{Imprecise Abstract Transfer Functions}
Widening operators are the best known source of non-monotonicity and incompleteness in finding suitable invariants in abstract interpretation, but they are not the only one.
There are, and this may be surprising, cases of non-monotonicity and incompleteness due to transfer functions, particularly in combinations of abstractions.

It is well-known that even if each individual step of abstract interpretation computes the least possible post-condition within the abstract domain compatible with the precondition, this property does not extend to composition of steps; in other words, optimality does not compose.
Let us see a simple example:
\begin{lstlisting}
y=x;
z=x-y;
\end{lstlisting}

If $x$ is known to lie in $[0,1]$, then so does $y$, and this is optimal---non-optimal but sound answers would be intervals containing $[0,1]$.
Then interval arithmetic deduces that $z$ is in $[-1,1]$, which is sound, but not optimal: the optimal interval is $[0,0]$.
However, in order to reach that conclusion, one must know that $x=y$ at the intermediate control point, that is, a relation between $x$ and $y$, which is not possible in a non-relational abstraction such as intervals.

One workaround to this weakness is to propagate a system of relations along with the interval analysis. For instance, we we had propagated $y=x$, then by substituting $y \mapsto x$ into $x-y$ and simplifying, we would have obtained $z=0$.

An approach to implement this workaround~\cite{Monniaux_ASIAN06}, for instance applied in the Astrée static analyzer, is to propagate a terminating rewriting system (for instance, populated in chronological order) along with the interval information, perform interval analysis on both the original expressions and the expressions obtained by rewriting and simplification, and then take the intersection.
Here, the rewriting system would contain $y \mapsto x$, and the rewritten and simplified expression would yield $z=0$.

This approach however creates non-monotonicity.%
\footnote{This non-monotonicity resulted in some ``false alarms'', that is, warnings about nonexistent problems, one some industrial programs.}
Consider this example~\cite[ex.~4]{Monniaux_ASIAN06}:
\begin{center}
\begin{tabular}{l|l|l}
Code & Symbolic computation & Less precise symbolic \\
\lstinline|int i, j=i+1;| & $j \mapsto i+1$ & nothing \\
  \lstinline|int k=j+1;| & $j \mapsto i+1$, $k \mapsto j+1 \mapsto i+2$ & $k \mapsto j+1$\\
\lstinline|if (j > 0) l=k;| & $j \mapsto i+1$, $k \mapsto i+2$, $l\mapsto i+2$ & $k \mapsto j+1$, $l \mapsto j+1$
\end{tabular}
\end{center}
Assume we don't know anything about the values of the variables initially, and consider the interval obtained for $l$ at the end of the ``then'' branch of the test.
Simple interval propagation yields no information about $l$.
The rewriting system yields $l \mapsto i+2$ and since no information is known about $i$, no information is known about~$l$.
Yet, if we forget that $j \mapsto i+1$, and apply the resulting rewriting system, we obtain $l \mapsto j+1$ and thus $l > 1$.

Granted, some improvements could be made by more clever propagation.%
\footnote{More generally, advanced tools such as Astrée implement complex combinations of domains and iteration strategies, which make many examples of non-monotone behavior and false alarms on simple analyses actually succeed.}
For instance, since $j \mapsto i+1$, the guard could be rewritten into $i +1 > 0$, and this could itself be rewritten into $i > -1$.
This amounts to ``inverting'' $j \mapsto i+1$ to $i \mapsto j-1$.
More generally, one would need to consider the rewriting system not as a directed system, but as a system of equations.
The combination of a system of linear equations and intervals forms the basis of the simplex algorithm for linear programming, and obtaining optimal interval bounds from a combination of equalities and interval bounds amounts to linear programming.
In fact, any convex polyhedron can be expressed as the projection of a set defined by the conjunction of linear equalities and interval bounds.%
\footnote{Consider the representation of the polyhedron as a system of inequalities $l_i(x_1,\dots,x_n) \leq B_i$, create a new variable $y_i$ for any linear combination $l_i$ for which there does not already exist a variable, keep this inequality as $y_i \leq B_i$ and $y_i = l_i(x_1,\dots,x_n)$, then the set defined by the original system is the projection of the set defined by the transformed system on the $x$ variables}
The appropriate domain for dealing with such constraints precisely would thus be that of convex polyhedra~\cite{CousotHalbwachs78,halbwachs:tel-00288805}.

\section{Undecidability of an Abstraction}
The argument from Section~\ref{sec:exact} for establishing decidability of the existence of an inductive invariant within the domain of convex polyhedra with at most $r$ faces does not extend to the domain convex polyhedra with any number of faces.
In fact, there is no known algorithm for deciding the existence of inductive invariants within that domain for any nontrivial class of programs.
Invariant inference approaches for that domain, starting from the early proposals from Cousot and Halbwachs~\cite{CousotHalbwachs78,halbwachs:tel-00288805}, are typically based on iterations with widening.
An intriguing question is whether it is inevitable to resort to heuristics.

\subsection{Polyhedral Abstraction}
I have attempted proving that there is no algorithm deciding the existence of polyhedral invariants for linear transition systems (real or integer), and have so far failed.
Because of my efforts in raising attention to this issue, some colleagues named this the ``Monniaux problem''.
Several distinguished researchers\footnote{Including some at the 2022 CSV workshop in Venice, whence this volume comes.}
have also reported trying to solve it and failing.

\subsubsection{Undecidability with Quadratic Guards}
Progress has however been made on variants of the so-called ``Monniaux problem''~\cite{DBLP:conf/sas/FijalkowLOOP019}.
I was able to prove that this problem becomes undecidable if one is allowed to use a quadratic transition guard: it is then possible to encode a deterministic counter machine reachability problem into the invariant inference problem. Let us recall how (proof sketch)~\cite{Monniaux_Acta_Informatica_2018}.

Let the counter machine $M$ operate over variables $z_1,\dots,z_n$, to which we add two variables $x$ and $y$, initialized to $0$. The transitions of the counter machine are synchronously combined with steps $(x,y) \mapsto(x+1,y+x)$.
The combined machine thus simulates the counter machine on variables $z_i$ together with a parabola $\big(n,\frac{1}{2}n(n-1)\big)$ on variables $(x,y)$, where $n$ is the number of steps taken so far.
Now modify the resulting machine by conjoining to all transitions the guard $y=\frac{1}{2}x(x-1)$; clearly this does not modify the behavior of the machine.
Add a special control state $\sigma_b$, meant to be unreachable, and transitions from any state $y<\frac{1}{2}x(x-1)$ to that state. Call the resulting machine~$M'$.

Assume $M$ terminates in $N$ steps, then so does $M'$. Consider the convex hull $H$ of the finite family of points such that $x=k$, $y=\frac{1}{2}k(k-1)$, $z_1,\dots,z_n$ is the state reached after $k$ steps in the execution of $M$, and $0 \leq k \leq N$.
The only points in $H$ that lie on the $y=\frac{1}{2}x(x-1)$ parabola project to the reachable states of~$M$.
$H$, in general, contains extra states (integer points) such that $y > \frac{1}{2}x(x-1)$, but due to the nonlinear guards, these states cannot transition to other states.
Thus $H$ is an inductive invariant for $M'$ that proves the inaccessibility of~$\sigma_b$.



\begin{figure}
  \begin{center}
  \begin{tikzpicture}
    \draw
(0, 0.0) --
(1, 0.0) --
(2, 0.5) --
(3, 1.5) --
(4, 3.0);
\draw[dotted]
(4, 3.0) --
(5, 4.5);
\draw[color=red]
(0.0, -0.0) --
(0.1, -0.022500000000000003) --
(0.2, -0.04000000000000001) --
(0.30000000000000004, -0.052500000000000005) --
(0.4, -0.06) --
(0.5, -0.0625) --
(0.6000000000000001, -0.06) --
(0.7000000000000001, -0.05249999999999999) --
(0.8, -0.039999999999999994) --
(0.9, -0.022499999999999996) --
(1.0, 0.0) --
(1.1, 0.027500000000000028) --
(1.2000000000000002, 0.06000000000000006) --
(1.3, 0.09750000000000002) --
(1.4000000000000001, 0.14000000000000007) --
(1.5, 0.1875) --
(1.6, 0.24000000000000005) --
(1.7000000000000002, 0.2975000000000001) --
(1.8, 0.36000000000000004) --
(1.9000000000000001, 0.4275000000000001) --
(2.0, 0.5) --
(2.1, 0.5775000000000001) --
(2.2, 0.6600000000000001) --
(2.3000000000000003, 0.7475000000000003) --
(2.4000000000000004, 0.8400000000000003) --
(2.5, 0.9375) --
(2.6, 1.04) --
(2.7, 1.1475000000000002) --
(2.8000000000000003, 1.2600000000000002) --
(2.9000000000000004, 1.3775000000000004) --
(3.0, 1.5) --
(3.1, 1.6275000000000002) --
(3.2, 1.7600000000000002) --
(3.3000000000000003, 1.8975000000000004) --
(3.4000000000000004, 2.0400000000000005) --
(3.5, 2.1875) --
(3.6, 2.3400000000000003) --
(3.7, 2.4975000000000005) --
(3.8000000000000003, 2.6600000000000006) --
(3.9000000000000004, 2.8275000000000006) --
(4.0, 3.0) --
(4.1000000000000005, 3.177500000000001) --
(4.2, 3.3600000000000003) --
(4.3, 3.5475) --
(4.4, 3.7400000000000007) --
(4.5, 3.9375) --
(4.6000000000000005, 4.1400000000000015) --
(4.7, 4.3475) --
(4.800000000000001, 4.560000000000001) --
(4.9, 4.777500000000001);
\end{tikzpicture}
\end{center}
\caption{When the process is non terminating, any polyhedron containing the reachable points inevitably goes below the parabola}
\label{fig:parabola}
\end{figure}
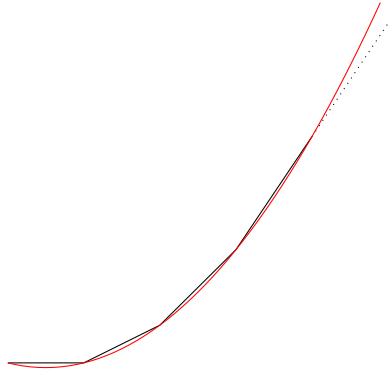

Assume $M$ does not terminate, and thus so does~$M'$.
Any inductive invariant for $M'$ must contain the infinite family of points  such that $x=k$, $y=\frac{1}{2}k(k-1)$, $z_1,\dots,z_n$ is the state reached after $k$ steps, for~$k \geq 0$.
Any convex polyhedron that contains this infinite family of points must have a point  $y < \frac{1}{2}x(x-1)$: assume there is none and consider the vertex $V^*$ with maximal $x^*$; this vertex must lie on the parabola; there is for $x \geq x^*$ at least one infinite face of the form $y \geq L(x, z_1, \dots, z_n)$; but then inevitably there will be points of that face lying below the parabola (Fig.~\ref{fig:parabola}).
However, any convex polyhedron that contains a point where $y < \frac{1}{2}x(x-1)$ is not an inductive invariant capable of proving that $\sigma_b$ is unreachable.
It follows that there is no convex polyhedron that is an inductive invariant for $M'$ capable of showing the unreachability of~$\sigma_b$.

Thus, $M'$ has an inductive polyhedral invariant suitable for proving the unreachability of~$\sigma_b$ if and only if $M$ terminates, and this for arbitrary $M$ in a class of machines with undecidable termination.

The same reasoning applies whether the state space is considered over integers or over the reals.

\subsubsection{Perspectives and Dead Ends}


The role of the parabola $y=\frac{1}{2}x(x-1)$ in the above proof could be played by other sets, such as a circle $x^2+y^2=1$: the process that enumerated points on the parabola can be replaced by one that enumerates points on the circle. Initialize $(x,y)=(1,0)$, and, as next step, apply a rotation matrix defined by a Pythagorean triple ($(a,b,c)$ integers such that $a^2+b^2=c^2$, for instance $(3,4,5)$):
$\left(\begin{smallmatrix}
  \frac{a}{c} & \frac{b}{c} \\
  \frac{b}{c} & -\frac{a}{c} \\
\end{smallmatrix}\right)$.
This however leads to more complex proof arguments, and does not gain anything: we still need a nonlinear guard.

Essentially, what we used in both cases is a strictly convex set $S$ ($y \geq \frac{1}{2}x(x-1)$) with boundary definable by a guard ($y=\frac{1}{2}x(x-1)$), with exterior ($y < \frac{1}{2}x(x-1)$) definable by a guard, and with a way to enumerate points in the boundary while still remaining in the class of deterministic transitions under consideration. Strict convexity is used so that taking the convex hull of points in $S$, as when using polyhedral invariants, does not add ``parasitic'' points on the boundary. This ensures that the guard that keeps only points on the boundary removes ``parasitic'' points.

Could we achieve similar goals using only linear constraints?
Suppose we can express such a guard with a formula, possibly containing disjunctions. Then, by distributivity, this formula expresses a finite union of polyhedra $P_1 \cup \dots \cup P_n$. Then there is a $i$ such that the enumeration process eventually picks two points (in fact, an infinity of them) in $P_i$; and thus the guard cannot exclude ``parasitic'' points between these two, at least over the reals.

A possible course of action could be to set the problem over integers and take advantage of the implicit guard that non-integer points are discarded.
This however was also a dead end.

\subsection{Richer Domains}
If the abstract domain (class of invariants) under consideration is rich enough to express exactly the set of reachable states of the programs to be analyzed, then obviously the problem is undecidable: a suitable inductive invariant (the set of reachable states) exists in the domain if and only if the safety property is true.

This happens for instance if the domain includes $\Sigma^0_1$ formulas of Peano arithmetic, that is, formulas of the form $\exists v_1 \dots \exists v_n~F$ where $F$ only contains bounded quantifiers and the usual arithmetic operations and comparisons.
By a form of Gödel's encoding, for any integer program with $m$ variables built using the usual arithmetic operators, it is possible to build a $\Sigma^0_1$ formula $F(k,s_1,\dots,s_m)$ satified if and only if $(s_1,\dots,s_m)$ is the state of the program after $k$ steps of~\cite[Ch.~7]{DBLP:books/daglib/0070910}\cite{doi:10.1137/0207005}.

Such domains are obviously too rich. In fact, with the domain of $\Sigma^0_1$ formulas described above, it is not even possible in general to check for inductiveness, because the formulas belong to an indecidable class.

However, even semilinear sets (sets defined by Boolean combinations of linear inequalities with algebraic coefficients, thus disjunctions of convex polyhedra) are sufficient to obtain undecidability even for a very restricted class of programs (one single loop control state, nondeterministic choice between two linear transformations, no guards)~\cite[Th.~9]{DBLP:conf/sas/FijalkowLOOP019}.

\section{Perspectives and Conclusion}
The question of the completeness of the domain with respect to the properties to prove, and the class of programs under consideration, is distinct from the problem of algorithmically finding suitable invariants within that domain---or the related problem of computing the least inductive invariant within the domain.

The traditional approach for abstract interpretation in domains with infinite ascending chains is iterations with widening. This approach has known weaknesses: the analysis may miss the invariants necessary for the proof and the analysis behavior is non-monotone (increasing knowledge about preconditions or transitions may decrease the precision of the analysis), which is a source of ``brittleness'' (a minor change in the program causes the analysis to fail to prove the property).
Many tricks are thus used to improve the invariants obtained with widening: narrowing iterations~\cite{DBLP:journals/logcom/CousotC92,Cousot78}, lookahead widening~\cite{DBLP:conf/cav/GopanR06}, guided static analysis~\cite{DBLP:conf/sas/GopanR07}. Though they help in practice, none of them guarantees that analysis will succeed.

Over time, completely different approaches were designed to compute suitable arguments for proving safety properties. While widening is a form of extrapolation (look at sets of reachable states in $0,1,2\dots$ steps and try to extrapolate for arbitrary number of steps), these methods are based on \emph{interpolation}: given some under-approximation of the set of reachable states, and the property to prove, find a simple separation between the two (a Craig interpolant) and hope that the interpolant, or components thereof, or formulas constructed from components of successive interpolants, becomes inductive.
One such approach is \emph{property-directed reachability} \cite{DBLP:conf/vmcai/Bradley11,DBLP:conf/fmcad/EenMB11}, later enriched with a number of extensions and improvements (dealing with arithmetic theories, dealing with non-linear Horn clauses and not just transition systems).
Variants of this approach are in particular implemented in the popular Z3 solver.%
\footnote{\url{https://github.com/Z3Prover/z3}}
Unfortunately, this approach, too, is brittle \cite{braine:hal-01337140}: minor differences in the problem to be solved (names of variables, etc.) may result in dramatic changes in outcome (finding invariants or timing out).

Approaches that are guaranteed to find the least inductive invariant in the chosen abstract domain (or, more generally, an inductive invariant suitable for proving a given property, if it exists) are more resilient;
I have listed several of them in Section~\ref{sec:exact}.
Some of these methods however do not scale up well (those based on quantifier elimination); policy iteration seems to be the most scalable.

In our view, the case for continued use of widening operators, despite their known weaknesses, or other non monotonic and/or brittle methods, would be strengthened by a proof that the existence of a suitable invariant in the domain is undecidable, or at least has high complexity.
Unfortunately, such undecidability proofs are hard.

In contrast, complete methods have many desirable properties. We have illustrated this with the example of LRU cache analysis (Section~\ref{sec:cache}). Since the end result has a unique definition, there is no brittleness, and several algorithms or implementations can be used to compute the same result, which allows performance comparisons (the meaning of performance differences between methods producing non comparable results is unclear) and validation by testing that results are identical.

\printbibliography
\end{document}